\documentclass[a4paper,11pt]{article}
\pdfoutput=1 

\usepackage[T1]{fontenc} 
\usepackage{amsmath,amssymb,mathtools}
\usepackage{color}
\usepackage{cite}
\usepackage[colorlinks=true,linkcolor=blueG, citecolor=redG, urlcolor=magentaG, bookmarks]{hyperref}
\usepackage{graphics}
\definecolor{comments}{RGB}{220,20,60}
\usepackage[text={17.1cm,24.6cm},centering]{geometry} 
\usepackage{braket}
\usepackage{comment}
\usepackage{tikz}

\usepackage{graphicx}
\usepackage{subcaption}
\usepackage{caption}
\usepackage{xcolor}
\usepackage{ifthen}

\newcommand{\nc}{\newcommand}
\nc{\ir}{\mathrm{i}}
\nc{\dd}{\mathrm{d}} 
\nc{\eE}{\mathrm{e}}
\nc{\Tr}{\text{Tr}}
\nc{\id}{\mathbb{I}}
\nc{\Z}{\mathcal{Z}}
\nc{\T}{\mathcal{T}}
\nc{\E}{\mathcal{E}}
\nc{\F}{\mathcal{F}}
\nc{\Om}{\Omega}
\nc{\N}{\mathcal{N}}
\nc{\spp}{\hspace{1pt}}
\nc{\spm}{\hspace{-1pt}}
\nc{\lb}{\label} 
\nc{\nn}{\nonumber} 
\nc{\ra}{\rangle} 
\nc{\la}{\langle} 
\nc{\Nn}{\mathcal{N}} 
\nc{\sigb}{\boldsymbol{\sigma_{\text{bd}}}}

\definecolor{blueG}{RGB}{51, 102, 204}
\definecolor{magentaG}{RGB}{214.2, 40.8, 132.6}
\definecolor{redG}{RGB}{229.5, 51., 102.}

\numberwithin{equation}{section}

\usepackage{authblk}

\begin{document}
\title{\bf Fermionic logarithmic negativity in the Krawtchouk chain}
\author[1,2]{Gabrielle Blanchet}
\author[1,2,3]{Gilles Parez}
\author[1,2,4]{Luc Vinet}
\affil[1]{\it Centre de Recherches Math\'ematiques, Universit\'e de Montr\'eal, P.O. Box 6128, Centre-ville
Station, Montr\'eal (Qu\'ebec), H3C 3J7, Canada}
\affil[2]{\it D\'epartement de Physique, Universit\'e de Montr\'eal, Montr\'eal (Qu\'ebec), H3C 3J7, Canada}
\affil[3]{\it Laboratoire d'Annecy-le-Vieux de Physique Th\'eorique LAPTh, CNRS,
Universit\'e Savoie Mont Blanc, 74000 Annecy, France}
\affil[4]{\it IVADO, 6666 Rue Saint-Urbain, Montr\'eal (Qu\'ebec), H2S 3H1, Canada}
\maketitle

\begin{abstract}
   The entanglement of non-complementary regions is investigated in an inhomogeneous free-fermion chain through the lens of the fermionic logarithmic negativity. Focus is on the Krawtchouk chain, whose relation to the eponymous orthogonal polynomials allows for exact diagonalization and analytical calculations of certain correlation functions. For adjacent regions, the negativity scaling corresponds to that of a conformal field theory with central charge $c=1$, in agreement with previous studies on bipartite entanglement in the Krawtchouk chain. For disjoint regions, we focus on the skeletal regime where each region reduces to a single site. This regime is sufficient to extract the leading behaviour at large distances. In the bulk, the negativity decays as $d^{-4 \Delta_f}$ with $\Delta_f=1/2$, where $d$ is the separation between the regions. This is in agreement with the homogeneous result of free Dirac fermions in one dimension. Surprisingly, when one site is close to the boundary, this exponent changes and depends on the parity of the boundary site $m=0,1,2,\dots$, with $\Delta_f^{\textrm{even}}=3/8$ and $\Delta_f^{\textrm{odd}}=5/8$. The results are supported by numerics and analytical calculations. 
\end{abstract}

\newpage

\tableofcontents

\section{Introduction}

Quantum entanglement plays a central role in the study of quantum many-body systems, both in and out of equilibrium \cite{amico2008entanglement,Laflorencie:2015eck}. In particular, various measures of entanglement display singular behaviour in the vicinity of quantum phase transitions, hence allowing for their detection and characterization \cite{OAFF02, ON02, VLRK03, CC04}. 

In the context of bipartite quantum systems $A \cup B$ in a pure state $|\psi\rangle$, the canonical entanglement measure is the celebrated \textit{entanglement entropy} $S_A$. It is defined as
\begin{equation}
    S_A =  -\Tr( \rho_A \log \rho_A), \qquad \rho_A=\Tr_B \big(|\psi\rangle\langle \psi|\big),
\end{equation}
where $\rho_A$ is the reduced density matrix of subsystem $A$. In one-dimensional quantum critical systems described by conformal field theories (CFT), the entanglement entropy of an interval of length $\ell$ embedded in an infinite system reads \cite{holzhey1994geometric,CC04}
\begin{equation}
    S_A = \frac{c}{3} \log \ell + a, 
\end{equation}
where $c$ is the central charge of the underlying CFT and $a$ is a non-universal constant. In stark contrast, the ground-state entanglement entropy of gapped Hamiltonians obeys an area law \cite{hastings2007area,wolf2008area,eisert2010colloquium}. 

In numerous situations, it is also relevant to quantify entanglement between subsystems which are not described by a pure state. For example, in a tripartite system $A_1\cup A_2\cup B$, the composite system $A_1 \cup A_2$ is described by a mixed reduced density matrix $\rho_{A_1,A_2}$. In this situation, the entanglement entropy $S_{A_1}$ does not properly quantify the entanglement between $A_1$ and $A_2$, because it also depends on the entanglement between $A_1$ and $B$. Instead, one uses the \textit{logarithmic negativity} \cite{VW02,plenio2005logarithmic}. This entanglement measure is based on the positive partial transpose (PPT) criterion \cite{P96}, which is a necessary condition for the density matrix $\rho_{A_1,A_2}$ to be separable, or not entangled. The logarithmic negativity, denoted $\E$, quantifies the violation of this criterion, and is defined as 
\begin{equation}\label{eq:Eb}
    \E = \log \big \lVert \rho_{A_1,A_2}^{T_1}\big \rVert_1.
\end{equation}
Here, $\lVert X \rVert_1 = \Tr \sqrt{X X^\dagger}$ is the trace norm, and $T_1$ indicates the partial transposition with respect to~$A_1$. 

For quantum systems with fermionic degrees of freedom, the notion of separability, and hence the definition of entanglement, is affected by the fermionic statistics \cite{banuls2007entanglement}. As a result, the standard (or bosonic) logarithmic negativity defined above has a number of shortcomings. Most notably, it fails to detect topological phases in the Kitaev chain \cite{SSR17} and vanishes for certain simple fermionic states which are manifestly not separable \cite{shapourian2019entanglement}. For fermionic systems, one thus uses the \textit{fermionic logarithmic negativity}~$\E_f$ \cite{SSR17,shapourian2019entanglement}. It is defined similarly to its bosonic counterpart of Eq.~\eqref{eq:Eb}, but where the partial transpose is replaced by the partial time-reversal operation. 

Similarly to the entanglement entropy, the logarithmic negativity allows one to detect and characterize quantum critical regimes. For one-dimensional quantum critical systems, both the bosonic and fermionic logarithmic negativity of two adjacent intervals of length $\ell_1$ and $\ell_2$ embedded in an infinite system scale as \cite{CCT12,CCT13,SSR17}
\begin{equation}
    \E_{(f)} = \frac{c}{4} \log \Big(\frac{\ell_1 \ell_2}{\ell_1+\ell_2}\Big)+b,
    \label{eq:ECFT}
\end{equation}
where $c$ is the central charge of the underlying CFT and $b$ is a constant. In contrast, the scaling of the logarithmic negativity as a function of the distance for disjoint subsystems strongly differs depending on whether the system has bosonic of fermionic degrees of freedom. For bosonic (or spin) systems, the logarithmic negativity decays faster than any power for CFTs in one \cite{CCT12,CCT13} and arbitrary dimensions \cite{parez2024entanglement}. For finite lattice models, it experiences an entanglement sudden death, i.e., a finite distance beyond which the subsystems become exactly disentangled. This sudden death of entanglement has notably been observed in the transverse-field Ising model \cite{OAFF02,JTBBJH18,parez2024entanglement} and resonating valence-bond (RVB) states \cite{parez2023separability}, indicating that, contrary to common beliefs, these systems do not possess long-range entanglement. In fermionic systems however, there is no sudden death of entanglement. For fermionic CFTs, the logarithmic negativity decays as a power law with the distance $d$ between the subsystems \cite{parez2024entanglement},
\begin{equation}
    \E_f = B \left( \frac{\ell_1 \ell_2}{d^2}\right)^{2 \Delta_f},
\end{equation}
for $d\gg \ell_{1,2}$, where $B$ is a constant and $\Delta_f$ is the smallest fermionic scaling dimension of the theory. This dichotomy between the entanglement scaling for bosons and fermions at large distances is understood in full generality within the \textit{fate of entanglement} picture \cite{parez2024fate}.

Free-fermion models play an important role in the study of entanglement in many-body systems~\cite{peschel2003calculation,fagotti2008evolution,peschel2009reduced,casini2009entanglement,rodriguez2009entanglement,song2012bipartite,eisler2013free,eisler2017analytical,bons,SRRC19,SR19,PBC21}. On the one hand, these models can be studied analytically and with exact numerical methods, and on the other they are relevant to describe and investigate realistic many-body phenomena. While most results regarding entanglement in free-fermion models pertain to spatially homogeneous systems, there is a growing interest for entanglement in inhomogeneous free-fermion systems \cite{eisler2009entanglement,DSVC17,Crampe:2019upj,Crampe:2021,crampe2020entanglement,FA20,FA21,parez2022multipartite,bernard2023entanglement,bernard2021entanglement,bernard2022entanglement,bernard2023computation,bernard2023absence,bernard2024entanglement,bonsignori2024entanglement}. However, results for the fermionic logarithmic negativity in inhomogeneous free-fermion models are still lacking in the literature. In this article, we initiate this important line of research by considering the negativity in the Krawtchouk chain \cite{Crampe:2019upj,FA21,bernard2022entanglement}. This model belongs to a larger family of inhomogeneous models based on orthogonal polynomials from the Askey scheme \cite{koekoek2010hypergeometric}.  

This paper is organized as follows. We review the definition and diagonalization of the Krawtchouk chain in Sec.~\ref{sec:Kraw}. In Sec.~\ref{sec:FLN}, we provide the definition of the fermionic logarithmic negativity, as well as its expression in term of the chopped correlation matrix for free-fermion models. We investigate the properties of the fermionic logarithmic negativity in the Krawtchouk chain for adjacent and disjoint regions in Secs.~\ref{sec:adj} and \ref{sec:dis}, respectively. We offer concluding remarks and perspectives in Sec.~\ref{sec:ccl}. Various analytical derivations are presented in App.~\ref{sec:app}.

\section{The Krawtchouk chain}\label{sec:Kraw}
\subsection{Definition and diagonalization}
The Krawtchouk model describes inhomogeneously coupled free fermions on a one-dimensional chain of length $N+1$ with open boundary conditions. By convention, the sites are labelled from $n=0$ to $n=N$. The Hamiltonian is 
\begin{equation}
H= \sum_{n=0}^{N-1} J_n\Big( c_{n+1}^\dag c_{n} +c_{n}^\dag c_{n+1}\Big) - \sum_{n=0}^{N} B_n c_n^\dag c_n
\label{eq:H}
\end{equation}
with 
\begin{equation}
   J_n = \sqrt{(N-n)(n+1)p(1-p)}, \qquad B_n = -(Np + n(1-2p))+\mu,
   \label{eq:JB}
\end{equation}
where $0 \leqslant p \leqslant 1$ is a parameter of the model and $\mu$ is the chemical potential. The fermionic creation and annihilation operators $c_{n}^{(\dagger)}$ satisfy the usual anticommutation relations.  

The diagonalization of the many-body Hamiltonian \eqref{eq:H} relies on the diagonalization of the single-particle Hamiltonian $\Lambda$. It is an $(N+1)\times (N+1)$ tridiagonal matrix, defined by $H = \sum_{ij} \Lambda_{ij}c_i^\dagger c_j$ with
\begin{equation}
    \Lambda = \begin{pmatrix}
        -B_0 & J_0 & 0 & \dots & 0 & 0  \\
        J_0 & -B_1 & J_1 & \dots & 0 & 0  \\
        0 & J_1 & -B_2 & \dots & 0 & 0  \\
        \vdots & \vdots & \vdots & \ddots & \vdots & \vdots  \\
       0 & 0 & 0  & \dots  & -B_{N-1} & J_{N-1}\\
      0 & 0 & 0 & \dots &   J_{N-1} & -B_{N}   \\
    \end{pmatrix}.
\end{equation}
The spectral problem reads 
\begin{equation}
    \Lambda v_k = \omega_k v_k, \quad k=0,1,\dots,N,
\end{equation}
and we write the components of the eigenvectors $v_k$ as $[v_k]_n \equiv \phi_k(n)$, where by convention we also have $n=0,1,\dots,N$. With the special choice for $J_n$ and $B_n$ in Eq.~\eqref{eq:JB}, this problem can be solved using Krawtchouk polynomials \cite{Crampe:2019upj}. More specifically, the eigenfunctions $\phi_k(n)$ and eigenvalues $\omega_k$ of the single-particle Hamiltonian are
\begin{equation}
\label{eq:phikn}
    \phi_k(n) = (-1)^n \sqrt{\Big(\frac{p}{1-p}\Big)^{n+k}(1-p)^N  \binom{N}{n} \binom{N}{k}}K_k(n,p,N), \qquad \omega_k = k-\mu,
\end{equation}
with $k=0,1,\dots,N$, where $K_k(n,p,N)$ are the Krawtchouk polynomials \cite{koekoek2010hypergeometric},
\begin{equation}
    K_k(n,p,N) = {}_2F_1\left(\begin{matrix}
        -n, \quad -k \\
        \quad -N \quad
    \end{matrix}; \frac{1}{p}\right).
\end{equation}

Under the canonical transformation
\begin{equation}
    d_k^{(\dag)} = \sum_{n=0}^N \phi_k(n) c_n^{(\dag)},
\end{equation}
the Krawtchouk Hamiltonian \eqref{eq:H} is recast in the following diagonal form,
\begin{equation}
    H = \sum_{k=0}^N \omega_k  d_k^{\dag} d_k,
\end{equation}
allowing for an immediate identification of its eigenvectors.
\subsection{Ground-state correlation functions}

We choose a chemical potential of the form 
\begin{equation}
    \mu = K +\epsilon, \quad 0<\epsilon<1,
\end{equation}
where $K>0$ is an integer, and $\epsilon$ is a small positive number which lifts the ground-state degeneracy. With this choice, the single-particle energies satisfy $\omega_K<0<\omega_{K+1}$, and therefore the ground state reads 
\begin{equation}
    |\Psi\rangle = d_0^{\dag}d_1^{\dag}\cdots d_K^{\dag} |0\rangle,
    \label{eq:GS}
\end{equation}
where $|0\rangle$ is the vacuum, with the property that $c_n|0\rangle = 0$ for all $n=0,1,\dots, N$. The filling fraction $\rho$ is defined as 
\begin{equation}
    \rho \equiv \frac{K+1}{N+1},
    \label{eq:rho}
\end{equation}
and we always consider a fixed, constant filling $\rho>0$ in the large-$N$ limit. 

The ground-state two-point correlation function is defined as $ C_{m,n} = \langle \Psi |c_m^\dag c_n |\Psi\rangle$. In terms of the single-particle eigenfunctions, it reads
\begin{equation}
\label{eq:CmnExact}
    C_{m,n} = \sum_{k=0}^K \phi_k(m) \phi_k(n)
\end{equation}
for $m,n=0,1,\dots, N$. For $m\neq n$, we use the Christoffel-Darboux formula (see, e.g., Ref.~\cite{andrews1999special}) and rewrite the correlation as
\begin{multline}
\label{eq:CmnCD}
    C_{m\neq n} = (-1)^{m+n}  \Big(\frac{p}{1-p}\Big)^{\frac{m+n}{2}+K+1}(1-p)^{N+1}\sqrt{\binom{N}{m}\binom{N}{n}}\binom{N}{K} (N-K)\\[.3cm] \times \frac{K_{K}(n,p,N)K_{K+1}(m,p,N)-K_{K}(m,p,N)K_{K+1}(n,p,N)}{n-m}.
\end{multline}



\section{Fermionic negativity in free-fermion models}\label{sec:FLN}

\begin{figure}
    \centering
\begin{center}
   \begin{tikzpicture}[scale=1.3]
\draw [very thick,blueG](0,0) -- (10,0);
\draw[ultra thick,redG](2,0) -- (4,0);
\draw[ultra thick,redG](6,0) -- (8,0);
\filldraw (0,0) circle (1.5pt);
\node[font=\Large] at (0,-0.25) {$0$};
\node[font=\Large] at (10,-0.25) {$N$};

\filldraw (2,0) circle (1.5pt);
\node[text=redG, font= \huge] at (3,0.6) {$A_1$};
\draw[<->,thick] (2,-0.45)--(4,-0.45);
\node[font=\Large] at (3,-0.65) { $\ell_1$};
\filldraw (4,0) circle (1.5pt);
\draw[<->, thick] (4,-0.45)--(6,-0.45);
\node[font=\Large] at (5,-0.65) {$d$};
\filldraw (6,0) circle (1.5pt);
\node [text=redG, font= \huge] at (7,0.6) {\textbf{$A_2$}};
\draw[<->,thick] (6,-0.45)--(8,-0.45);
\node[font=\Large] at (7,-0.65) {$\ell_2$};
\filldraw (8,0) circle (1.5pt);
\filldraw (10,0) circle (1.5pt);
\end{tikzpicture} 
\end{center}
    \caption{Geometry of a tripartite open chain.}
    \label{fig:schema chaine}
\end{figure}
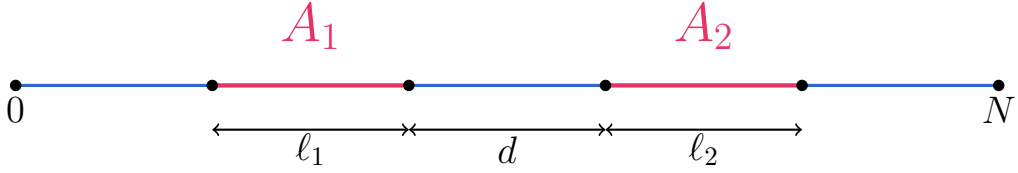

In this section, we review how the fermionic logarithmic negativity is calculated in free-fermion models. The goal is to investigate the entanglement between two intervals $A_1$ and $A_2$, of respective lengths $\ell_1$ and $\ell_2$, embedded in an open chain of length $N+1$ and separated by a distance $d$. This geometry is illustrated in Fig.~\ref{fig:schema chaine}. Moreover, we consider the case where the whole system is in the ground state $|\Psi\rangle$, defined in Eq.~\eqref{eq:GS}. We introduce the reduced density matrix $\rho_{A_1,A_2}$ of the composite system~$A_1 \cup A_2$,
\begin{equation}
   \rho_{A_1,A_2} = \Tr_B (|\Psi\rangle \langle \Psi|),
\end{equation}
where $B$ is the complement of $A_1\cup A_2$. The fermionic logarithmic negativity is defined as \cite{SSR17}
\begin{equation}
 \E_f = \log \big \lVert \rho_{A_1,A_2}^{R_1}\big \rVert_1,
\end{equation}
where $\lVert X \rVert_1 = \Tr \sqrt{X X^\dagger}$ is the trace norm, and $R_1$ indicates the partial time reversal on~$A_1$. This operation is similar to the standard partial transposition, used in the definition of the bosonic logarithmic negativity, see Eq.~\eqref{eq:Eb}, but adds a phase which depends on the fermion number in subsystems $A_1$ and~$A_2$. In the occupation basis, the partial time reversal operation is defined as follows. Consider basis states $|\alpha_1\rangle$ and $|\beta_2\rangle$ for the system $A_1$ and $A_2$ of the form 
\begin{equation}
    |\alpha_1\rangle = \prod_{j\in A_1} (c_j^\dagger)^{n_j} |0\rangle, \quad  |\beta_2\rangle = \prod_{j\in A_2} (c_j^\dagger)^{n_j} |0\rangle, 
\end{equation}
with occupation numbers $f_1 = \sum_{j\in A_1}n_j$ and $f_2 = \sum_{j\in A_2}n_j$, and $n_j=0,1$. 
The partial time reversal operation is then defined as \cite{SSR17}
\begin{equation}
    \Big( |\alpha_1\rangle|\beta_2\rangle\, \langle\widetilde\alpha_1|\langle\widetilde\beta_2|\Big)^{R_1}=(-1)^{\phi(\{n_j\},\{\tilde n_j\})}|\widetilde\alpha_1\rangle|\beta_2\rangle\, \langle\alpha_1|\langle\widetilde\beta_2|
\end{equation}
where the phase is 
\begin{equation}
   \phi(\{n_j\},\{\tilde n_j\}) = \frac{f_1(f_1+2)}{2}+ \frac{\tilde f_1(\tilde f_1+2)}{2} + f_2 \tilde f_2 + f_1 f_2 + \tilde f_1 \tilde f_2 + (f_1+f_2)(\tilde f_1 + \tilde f_2). 
\end{equation}



In free-fermion systems, Wick theorem implies that all correlation functions can be given in terms of two-point correlation functions. In turn, the reduced density matrix of a subsystem $A$, and hence entanglement measures, can be expressed in terms of the correlation matrix restricted to subsystem $A$, denoted~$\mathcal{C}_A$ \cite{peschel2003calculation}.
For a bipartite subsystem $A=A_1 \cup A_2$, this matrix admits a block structure, 
%
%
\begin{equation}
    \mathcal{C}_A = \begin{pmatrix}
        \mathcal{C}_{11} & \mathcal{C}_{12} \\
        \mathcal{C}_{21} &  \mathcal{C}_{22}\\
    \end{pmatrix},
    \label{eq:CorrMat Block}
\end{equation}
where the blocks $\mathcal{C}_{ij}$, of dimension $\ell_i \times \ell_j$, contain the correlations between the regions $A_i$ and $A_j$. The entries of these blocks are the two-point correlation functions $C_{m,n}$ given in \eqref{eq:CmnExact} (or equivalently~\eqref{eq:CmnCD} for the off-diagonal terms $m\neq n$), where the indices $m$ and $n$ are restricted to the sites in $A_i$ and $A_j$, respectively. For simplicity, we introduce the covariance matrix $J_{A} = 2 \mathcal{C}_A - \mathbb{I}$, which also has a block structure,
\begin{equation}
        J_{A} = \begin{pmatrix}
        J_{11} & J_{12} \\
        J_{21} & J_{22}\\
    \end{pmatrix}.
    \label{eq:J Block}
\end{equation}

The partial time-reversal operation modifies the covariance matrix. In particular, the matrices associated with $\rho_{A_1,A_2}^{R_1}$ and $(\rho_{A_1,A_2}^{R_1})^\dagger$, and denoted $J_+$ and $J_-$, respectively, are given by \cite{SR19,SRRC19} 
\begin{equation}
    J_{\pm} = \begin{pmatrix}
        -J_{11} & \pm \textrm{i} J_{12} \\
        \pm \textrm{i} J_{21} & J_{22}\\
    \end{pmatrix}.
    \label{eq:Jpm}
\end{equation}
Finally, let us define the matrix $J_{\rm x}$ as
\begin{equation}
    J_{\rm x} = (\mathbb{I} + J_+ J_-)^{-1} (J_+ + J_-).
\end{equation}
Using manipulations of Gaussian operators \cite{fagotti2010entanglement,eisert2018entanglement}, one can show that $J_{\rm x}$ is the covariance matrix corresponding to the product $\rho_{A_1,A_2}^{R_1} (\rho_{A_1,A_2}^{R_1})^\dagger$.
%
%
The fermionic logarithmic negativity is then \cite{SR19,SRRC19,murciano2021symmetry,parez2022dynamics}
\begin{equation}
    \mathcal{E}_f = \sum_{j=1}^{\ell_1 + \ell_2} \log \left[ \left( \frac{1+\nu_j^{\rm x}}{2} \right)^{\frac{1}{2}} + \left( \frac{1-\nu_j^{\rm x}}{2} \right)^{\frac{1}{2}} \right] + \frac{1}{2} \sum_{j=1}^{\ell_1 + \ell_2} \log \left[ \left( \frac{1+\nu_j}{2} \right)^2 + \left( 
 \frac{1-\nu_j}{2}\right)^2 \right],
\label{eq:LN}
\end{equation}
where $\nu_j^{\rm x}$ and $\nu_j$ are the eigenvalues of the matrices $J_{\rm x}$ and $J_A$, respectively.



\section{Adjacent regions}\label{sec:adj}








\begin{figure}
    \centering
    \includegraphics[width=0.49\linewidth]{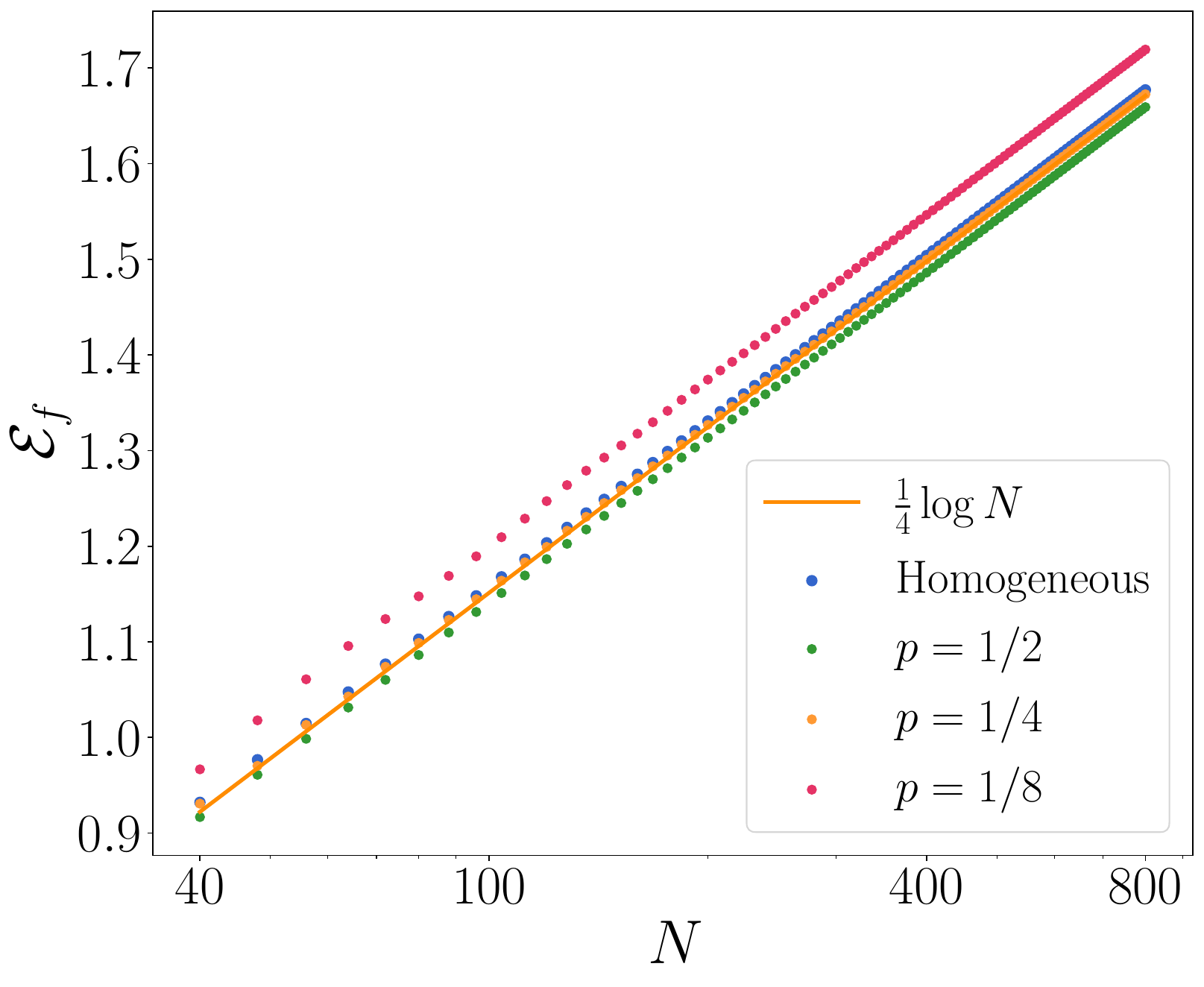}
   \includegraphics[width=0.49\linewidth]{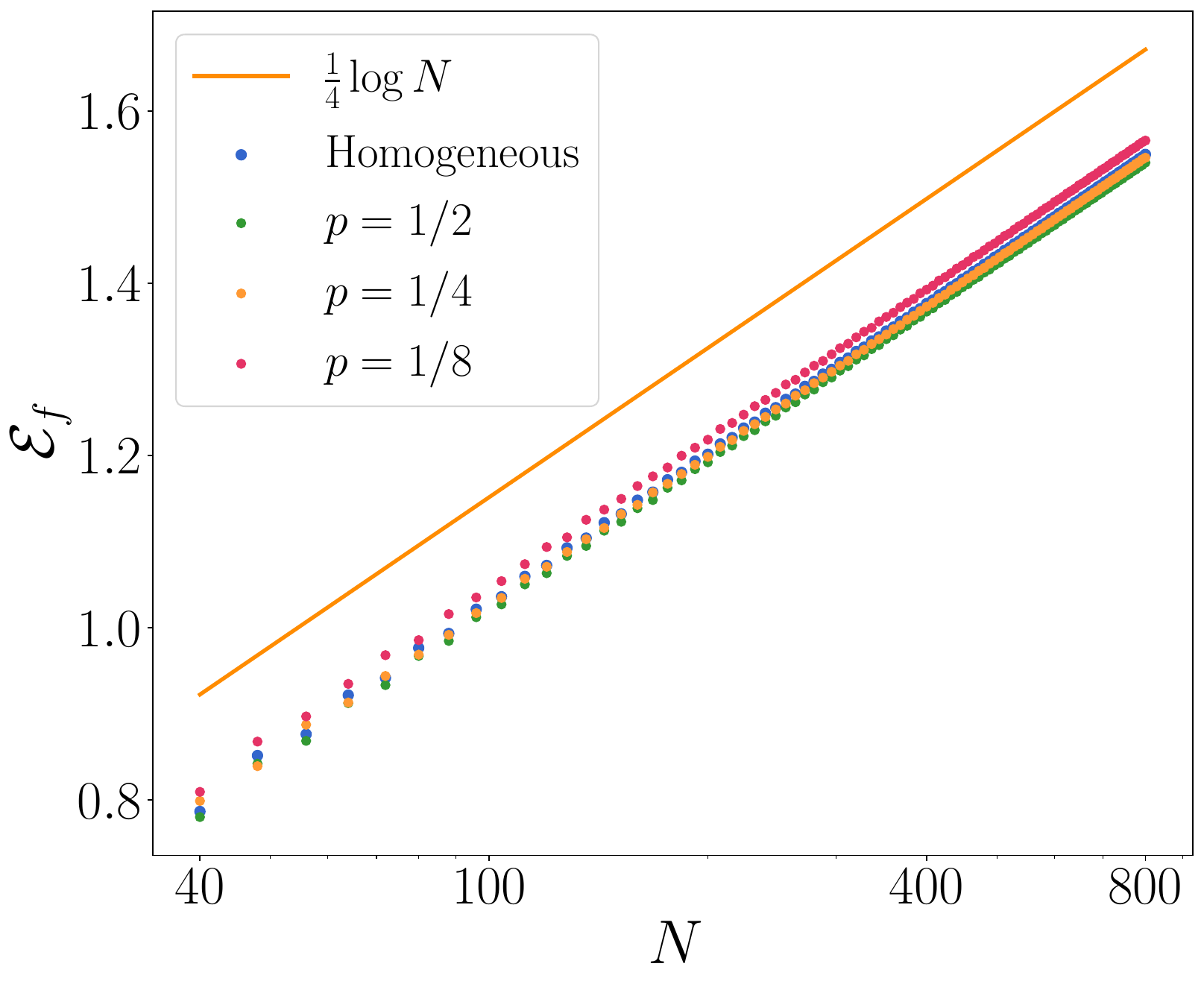}
    \caption{Fermionic logarithmic negativity $\E_f$ in log-scale for adjacent regions centered in the middle of the chain, of length $\ell_1=\ell_2 = \frac{N+1}{4}$ (left) and $\ell_1=2\ell_2 = \frac{N+1}{4}$ (right). $\E_f$ is plotted as a function of $N$ for the homogeneous chain and the Krawtchouk chain for various values of $p$. The symbols are obtained by numerical evaluation of Eq.~\eqref{eq:LN}, and the solid line is the curve $\frac 14 \log N$, which serves as a guide to the eye. Clearly all curves are parallel, indicating the same leading term $\E_f \sim \frac 14 \log N$.}
    \label{fig:graphs_adj}
\end{figure}

We start our investigations with the case of adjacent regions. For simplicity, we impose that the regions are centered in the middle of the chain, namely their contact point is located at site $(N+1)/2$. This geometry corresponds to Fig.~\ref{fig:schema chaine} with $d=0$ and where the right end of system $A_1$ is in the center of the chain. 
We then consider two cases: (i) $\ell_1=\ell_2=\frac{N+1}{4}$ and (ii) $\ell_1=2\ell_2=\frac{N+1}{4}$. In both cases, we study the scaling of the negativity as a function of the system size $N$ for the Krawtchouk chain at half filling $\rho=1/2$ with various values of $p$, as well as for the homogeneous free-fermion chain obtained by choosing $J_n=- \frac 12$ and $B_n=0$ in the Hamiltonian~\eqref{eq:H}. The diagonalization of the latter is standard, see, e.g., Ref.~\cite{Crampe:2019upj}. We report our numerical results in Fig.~\ref{fig:graphs_adj}. In all cases, we find a scaling of the form 
\begin{equation}
    \E_f = \frac{1}{4}\log N + \textrm{cst},
\end{equation}
which corresponds to the CFT scaling of Eq.~\eqref{eq:ECFT} with $c=1$. This value of the central charge is consistent with previous results regarding bipartite entanglement in the Krawtchouk chain \cite{FA21,bernard2022entanglement}, and it is a well-known fact for the homogeneous chain, see, e.g., Ref~\cite{VLRK03}. For the case (i), Eq.~\eqref{eq:ECFT} further predicts a constant term of the form $-\frac 14 \log 8 +b$, whereas for case (ii) it is  $-\frac{1}{4} \log 12 +b$. In all cases we find $b \sim 0.5$. A refined curved-space CFT \cite{DSVC17} analysis of the Krawtchouk chain is needed to properly interpret the physical content of the constant and subleading terms, and we leave this for future investigations.



\section{Disjoint regions}\label{sec:dis}

In this section, we investigate the decay of the fermionic logarithmic negativity as a function of the distance in the Krawtchouk chain. In this context, we consider the skeletal case \cite{BWK22} where both regions consist of a single site, $\ell_1=\ell_2=1$. This skeletal regime faithfully captures the leading order of the scaling of the negativity. Indeed, at large distances the two systems become point-like compared to their separation \cite{parez2024entanglement}. This approach was notably used to characterize the decay of the negativity for Dirac fermions in arbitrary dimension \cite{parez2024entanglement} and for the Schwinger model \cite{florio2024two}. 

For a state on which the fermion number operator is diagonal, the skeletal fermionic logarithmic negativity can be expressed in closed form as a function of the filling fraction $\rho$, the two-point correlation function $C_{m,n}$ and the density-density correlation $R_{m,n} = \langle \Psi|c_m^\dagger c_m c_n^\dagger c_n |\Psi\rangle$. For half-filling $\rho=1/2$, it reads \cite{parez2024entanglement}
\begin{equation}
    \E_f = \log \Big( 1-2R_{m,n}+2\sqrt{C_{m,n}^2+R_{m,n}^2} \Big). 
\end{equation}
The generalization to arbitrary filling is straightforward. 
In the limit of large separation, we have $\lim_{|m-n|\to \infty}C_{m,n} = 0$ and $\lim_{|m-n|\to \infty}R_{m,n} = \rho^2$. We expand the logarithmic negativity for arbitrary~$\rho$ in this limit and find
\begin{equation}
    \E_f = \frac{2}{1+2(\rho-1)\rho} |C_{m,n}|^2 + \mathcal{O}\Big(|C_{m,n}|^2 (R_{m,n}-\rho^2)\Big).
    \label{eq:EfC}
\end{equation}
For $\rho=1/2$, we recover the leading term $\E_f = 4 |C_{m,n}|^2$ \cite{parez2024entanglement}. 

\subsection{Bulk negativity}

\begin{figure}
    \centering
\begin{center}
   \begin{tikzpicture}[scale=1.3]
\draw [very thick,blueG](0,0) -- (10,0);

\node[font=\Large] at (0,-0.275) {$0$};
\node[font=\Large] at (6,-0.275) {$pN$};
\node[font=\Large] at (10,-0.275) {$N$};

\filldraw (0,0) circle (1.5pt);
\filldraw (6,0) circle (1.5pt);
\filldraw (10,0) circle (1.5pt);
\filldraw[redG] (4.5,0) circle (2.5pt);
\node[font=\Large] at (4.5,-0.275) {$m$};
\filldraw[redG] (7.5,0) circle (2.5pt);
\node[font=\Large] at (7.5,-0.275) {$n$};

\draw[<->, thick] (4.5,0.45)--(6,0.45);
\node[font=\Large] at (5.25,0.65) {$d/2$};
\draw[<->, thick] (6,0.45)--(7.5,0.45);
\node[font=\Large] at (6.75,0.65) {$d/2$};

\end{tikzpicture} 
\end{center}
    \caption{Illustration of the geometry for the bulk negativity. The two sites $m$ and $n$ are separated by a distance $d$ and centered around the position $pN$. }
    \label{fig:bulkCmn}
\end{figure}
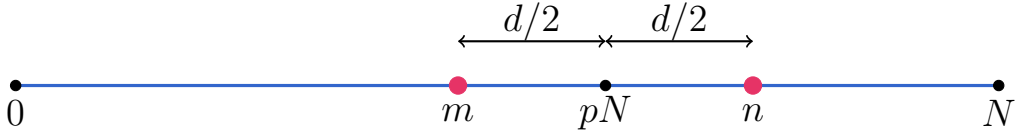
We consider the skeletal negativity between single sites as a function of the separation $d$ when the two sites are located well within the bulk of the chain. In particular, we choose to center them around the position $pN$, where $0<p<1$ is the parameter of the Krawtchouk chain (we exclude the extreme cases $p=0$ and $p=1$). This choice allows to perform analytical calculations for the two-point correlation function and hence for the negativity, with Eq.~\eqref{eq:EfC}. We depict this geometry in Fig.~\ref{fig:bulkCmn}. In the following, we work in the limit of large system size $N \gg 1$ and large separation $d\gg1$, with $d\ll N$.

\begin{figure}
    \centering
     \includegraphics[width=0.51\linewidth]{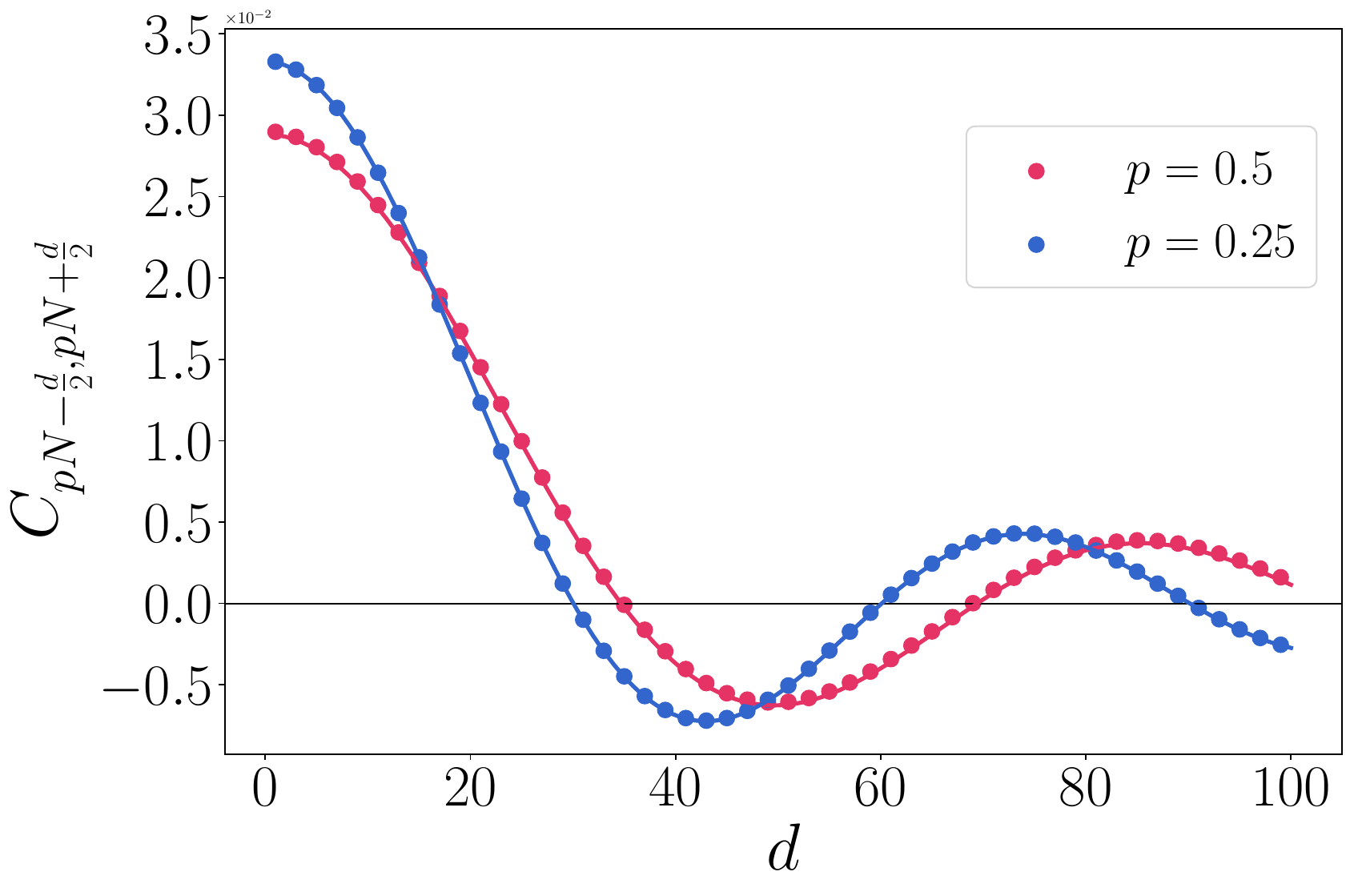}
    \includegraphics[width=0.475\linewidth]{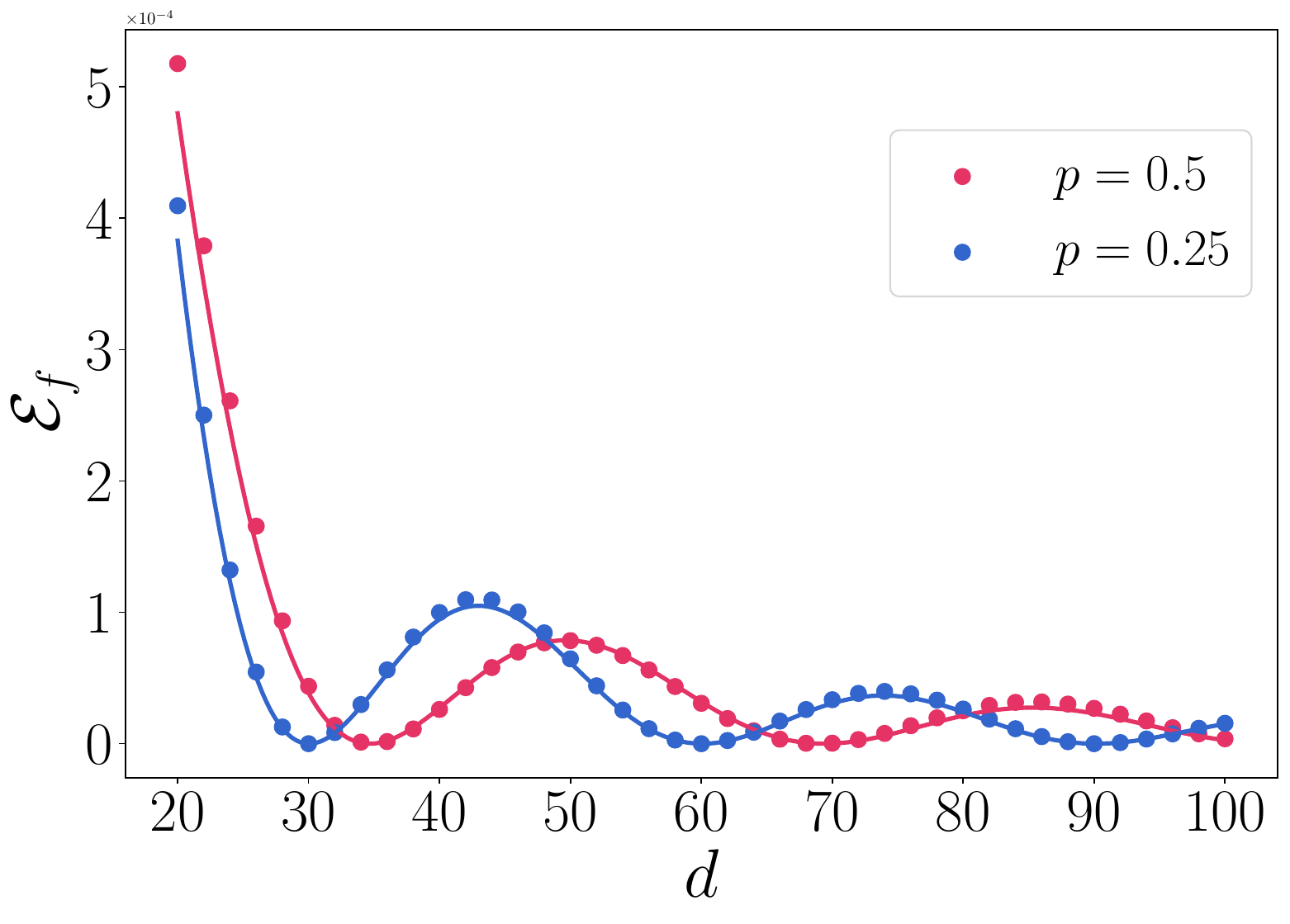}
    \caption{Two-point correlation function (left) and fermionic logarithmic negativity (right) between two sites centered around $pN$ as a function of the distance~$d$ with $N=2\times 10^4$ and small filling fraction $\rho = 2 \times 10^{-3}$ for various values of $p$. \textit{Left:} The symbols are obtained by numerical evaluation of Eq.~\eqref{eq:CmnExact}. The solid lines represent the analytical prediction of Eq.~\eqref{eq:CSmallFFAnalytical} for small filling fraction~$\rho$. \textit{Right:} The symbols are obtained by numerical evaluation of Eq.~\eqref{eq:LN}. The solid lines correspond to Eq.~\eqref{eq:EfC} where the two-point correlation function is given by Eq.~\eqref{eq:CSmallFFAnalytical}. The matches are extremely convincing.  }
    \label{fig:EfBulkSmallFF}
\end{figure}

First, we consider the case of small filling fraction $\rho \ll 1$. In this limit, we find
\begin{equation}
    C_{pN-\frac d2,pN+\frac d2 } \sim \frac{1}{\pi d}\sin\left( d \sqrt{\frac{\rho}{p(1-p)}} \right). 
    \label{eq:CSmallFFAnalytical}
\end{equation}
The proof of this result is provided in App.~\ref{sec:bulkC}. In Fig.~\ref{fig:EfBulkSmallFF}, we test this analytical prediction for the two-point correlation function, and the corresponding leading term for the logarithmic negativity obtained by combining Eqs.~\eqref{eq:EfC} and \eqref{eq:CSmallFFAnalytical}. We find an excellent match between the analytical prediction and the exact numerical calculations for both quantities.

For the interesting case of arbitrary filling fraction $\rho$, it is also possible to use known results regarding the asymptotics of Krawtchouk polynomials at $p=1/2$, see Ref.~\cite{ismail1998strong}, to extract the behaviour of the two-point correlation at large distance. We find
\begin{equation}
    C_{\frac{N-d}{2},\frac{N+d}{2}} \sim \frac{1}{\pi d} \sin \left( 2 d \arcsin (\sqrt{\rho})\right),
    \label{eq:CRhoPhalf}
\end{equation}
in the limit $N\to \infty$ with even $N$, and present the proof in App.~\ref{sec:Cphalf}.

\begin{figure}
    \centering
    \includegraphics[width=0.65\linewidth]{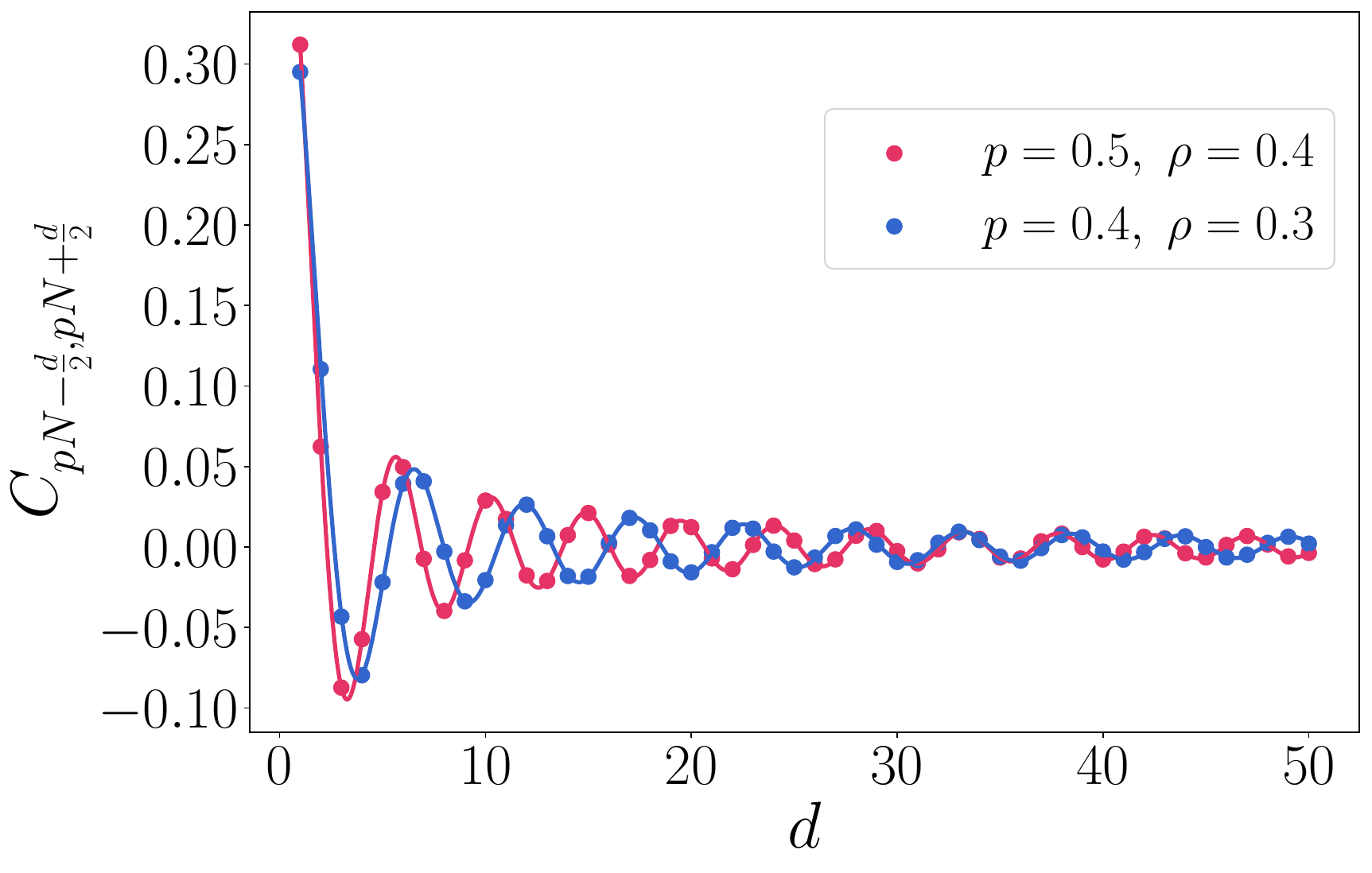}\\
    \includegraphics[width=0.475\linewidth]{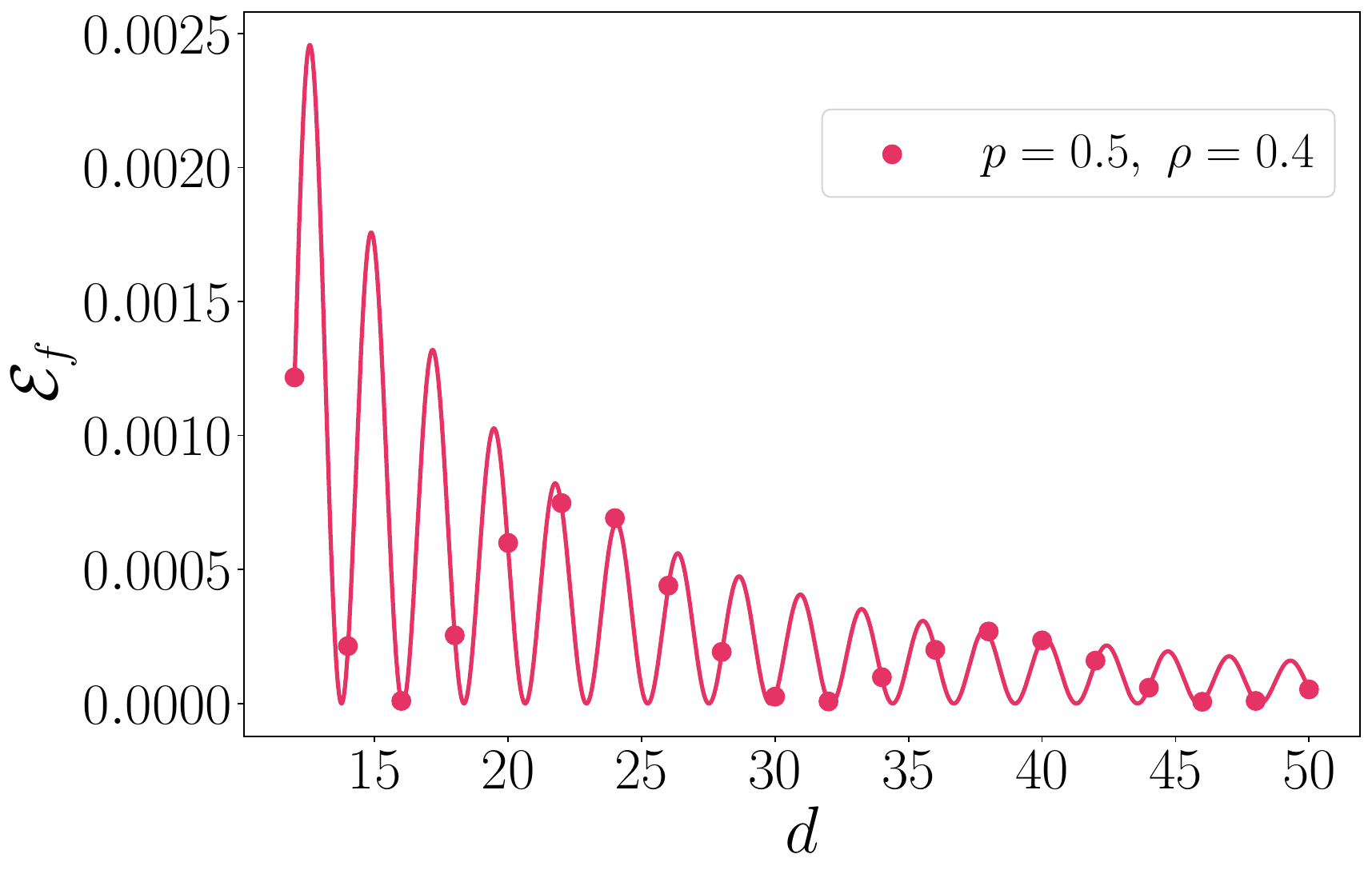}
     \includegraphics[width=0.475\linewidth]{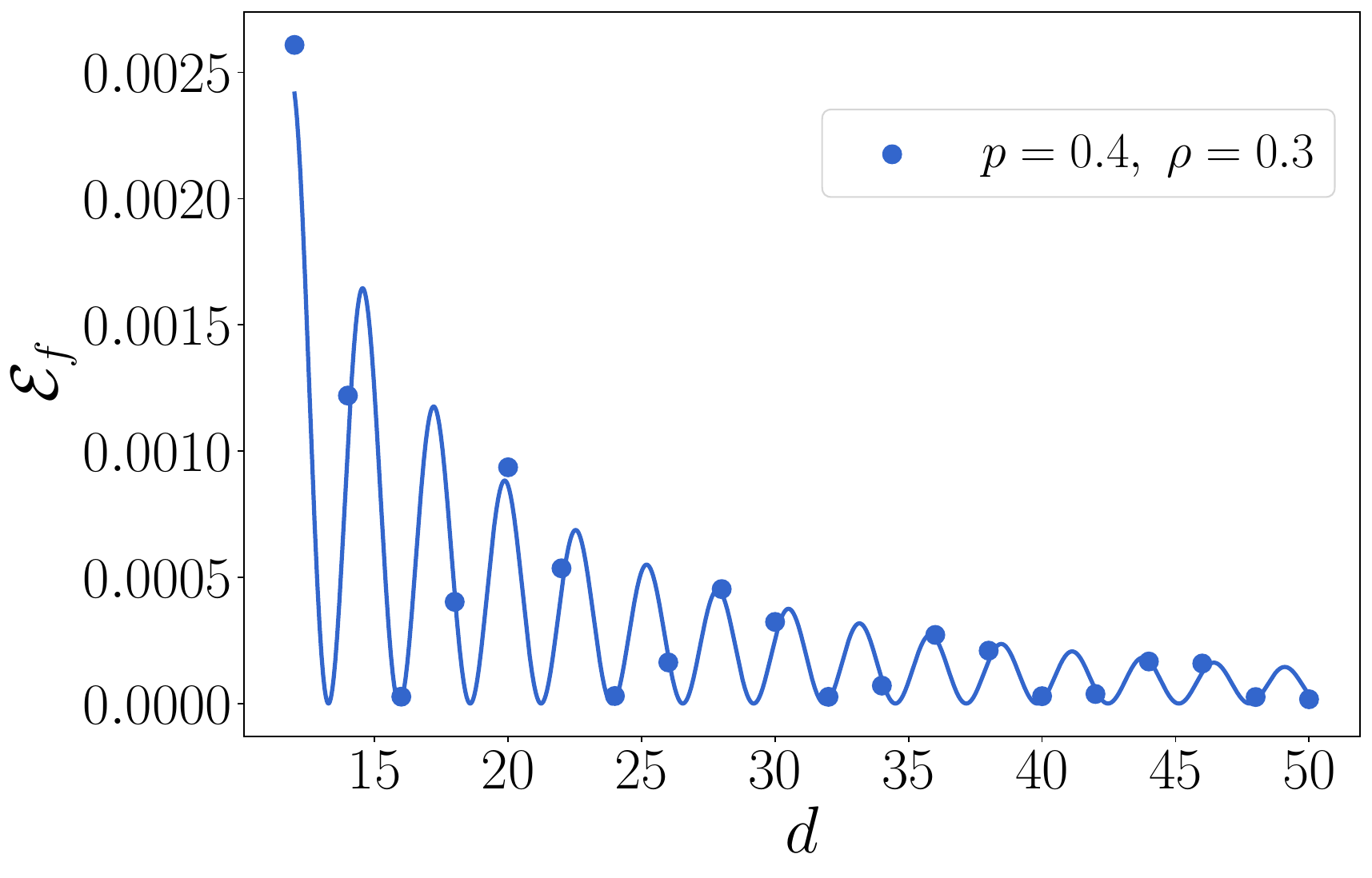}
    \caption{Two-point correlation function (top) and fermionic logarithmic negativity (bottom) between two sites centered around $pN$ as a function of the distance~$d$ with $N=4000$ for various values of $\rho$ and~$p$. \textit{Top:} The symbols are obtained by numerical evaluation of Eq.~\eqref{eq:CmnExact}. The solid lines represent the conjecture of Eq.~\eqref{eq:CGenCOnj} (or equivalently Eq.~\eqref{eq:CRhoPhalf} for $\rho=1/2$). \textit{Bottom:} The symbols are obtained by numerical evaluation of Eq.~\eqref{eq:LN}. The solid lines correspond to Eq.~\eqref{eq:EfC} where the two-point correlation function is given by Eq.~\eqref{eq:CGenCOnj}.}
    \label{fig:Efgen}
\end{figure}

Comparing Eqs.~\eqref{eq:CSmallFFAnalytical} and \eqref{eq:CRhoPhalf}, we conjecture that the two-point correlation function scales as 
\begin{equation}
    C_{pN-\frac d2,pN+\frac d2 } \sim \frac{1}{\pi d}\sin\left( d \sqrt{\frac{1}{p(1-p)}}  \arcsin (\sqrt{\rho})\right)
    \label{eq:CGenCOnj}
\end{equation}
for arbitrary $\rho$ and $p$ in the regime $1\ll d \ll N$. 
In Fig.~\ref{fig:Efgen}, we compare the conjecture of Eq.~\eqref{eq:CGenCOnj} with numerical calculations and find an excellent agreement, both for the two-point function and the logarithmic negativity. In the numerical calculations, we take $N=4000$ and choose a simple rational number for the filling fraction. We then take $K = \lceil \rho(N+1)-1\rceil$, which is equivalent to the definition of Eq.~\eqref{eq:rho} in the large-$N$ limit.

Disregarding the oscillatory terms, we thus find with Eqs.~\eqref{eq:EfC} and \eqref{eq:CGenCOnj} that the bulk negativity in the Krawtchouk chain decays as $d^{-2}$:
\begin{equation}
    \E_f \sim \frac{2}{(1+2(\rho-1)\rho)\pi^2} \frac{1}{d^2}. 
    \label{eq:EfbulkScaling}
\end{equation}
This is exactly the same behaviour as in the homogeneous chain. Indeed, in the uniform model, the correlation function in the infinite open chain reads \cite{peschel2009reduced,fagotti2011universal}
\begin{equation}
  C^{\textrm{hom}}_{m,n} = \frac{\sin (\pi \rho (m-n))}{\pi (m-n)} - \frac{\sin (\pi \rho (m+n))}{\pi (m+n)}.
    \label{eq:CHom}
\end{equation}
Deep in the bulk, the second term in Eq.~\eqref{eq:CHom} vanishes, and from Eq.~\eqref{eq:EfC} we get $\E_f^{\textrm{hom}} \sim \frac{2}{(1+2(\rho-1)\rho)\pi^2} \frac{1}{d^2}$, as in Eq.~\eqref{eq:EfbulkScaling}. As expected, this is also the same power-law decay as for Dirac fermions in one dimension \cite{parez2024entanglement}.

\subsection{Boundary negativity}

\begin{figure}
    \centering
\begin{center}
   \begin{tikzpicture}[scale=1.3]
\draw [very thick,blueG](0,0) -- (10,0);

\node[font=\Large] at (0,-0.275) {$m=0$};
\node[font=\Large] at (10,-0.25) {$N$};
\filldraw[redG] (0,0) circle (2.5pt);
\filldraw (10,0) circle (1.5pt);

\filldraw[redG] (4,0) circle (2.5pt);
\node[font=\Large] at (4,-0.275) {$n$};

\draw[<->, thick] (0,0.45)--(4,0.45);
\node[font=\Large] at (2,0.65) {$d$};

\end{tikzpicture} 
\end{center}
    \caption{Illustration of the geometry for the boundary negativity. The two sites $m$ and $n$ are separated by a distance $d$, and $m=0$ is at the left end of the chain. We also consider the case $m=1$, but the figure is very similar.}
    \label{fig:bndrCmn}
\end{figure}
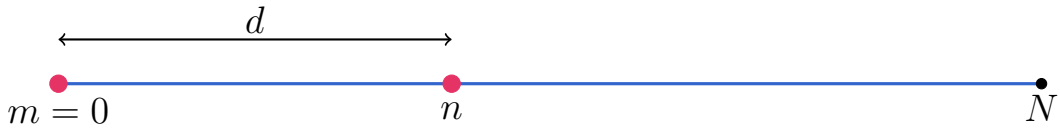

\begin{figure}
    \centering
    \includegraphics[width=0.7\linewidth]{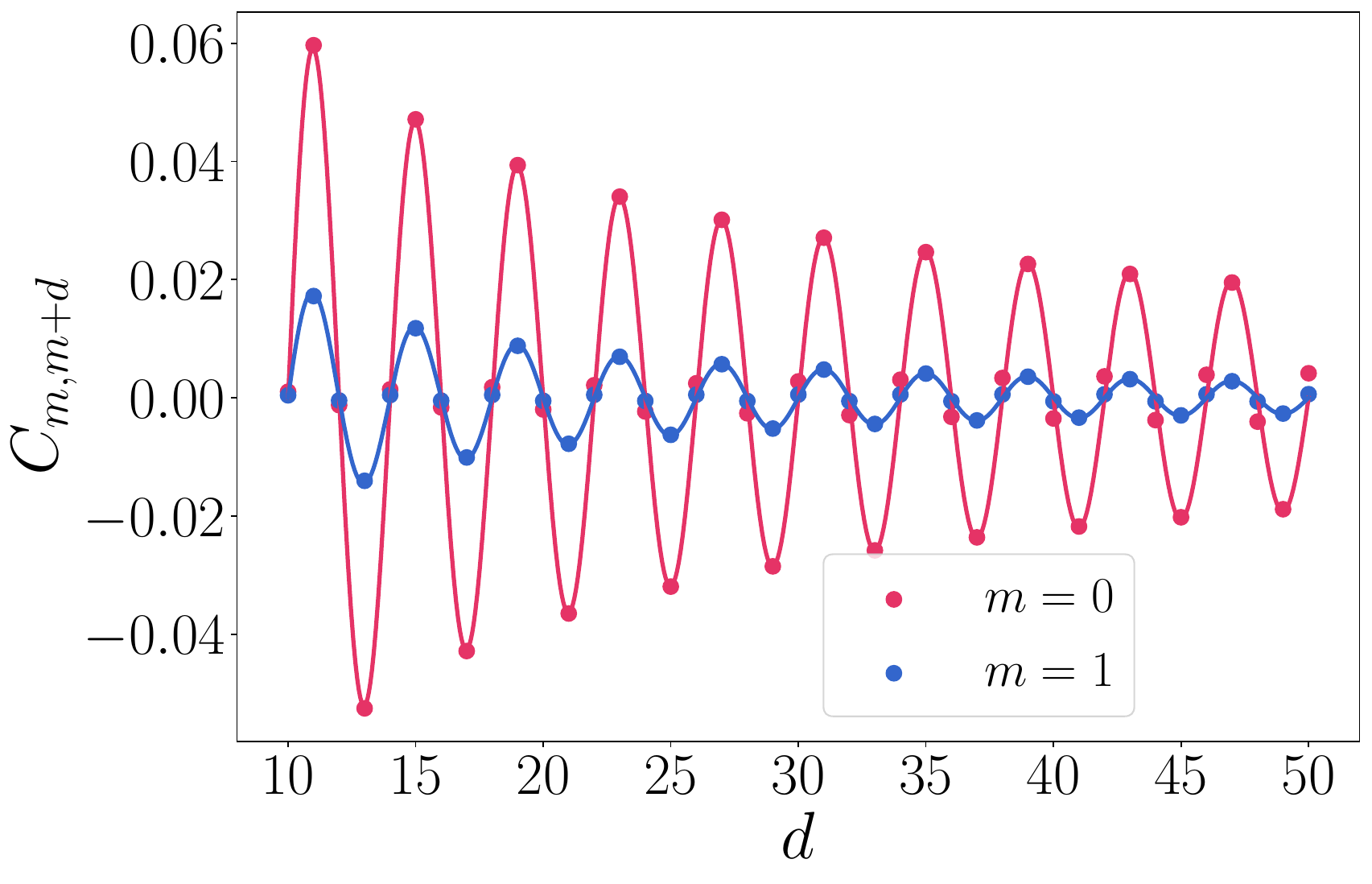}\\
    \caption{Boundary correlation $C_{m,m+d}$ with $m=0,1$ for $N=10^{6}$ and $\rho=p=0.15$. The symbols are obtained by numerical evaluation of the correlation functions, and the solid lines are the analytical predictions of Eq.~\eqref{eq:C01d}. We find an excellent agreement.}
    \label{fig:Cbndr}
\end{figure}

Because the Krawtchouk chain is inhomogeneous, it is important to investigate the scaling of the negativity in different regions of the chain. We thus turn to the examination of boundary negativity, where we study the negativity between the first or second site of the chain and a site at distance $d$. We illustrate this geometry in Fig.~\ref{fig:bndrCmn}. We still work in the regime $1\ll d \ll N$. For the special case where $\rho=p$ in the large-$N$ limit, we find
\begin{subequations}
\label{eq:C01d}
\begin{align}
    C_{0,d} & \sim\frac{1}{(2\pi^3)^{\frac 14} d^{\frac 34}}\sin\Big(\frac{-\pi d}{2}\Big), \label{eq:C0d} \\
    C_{1,d+1} & \sim \frac{1}{(2\pi^3)^{\frac 14} d^{\frac 54}}\sin\Big(\frac{-\pi d}{2}\Big). \label{eq:C1dplus1}
\end{align}
\end{subequations}
Choosing to set $\rho=p$ allows to derive these results analytically. The proof is provided in App.~\ref{sec:bndrC}. We test these predictions against numerical evaluations of the correlation functions in Fig.~\ref{fig:Cbndr}. We note that, in order to find a good match between the numerics and the analytical predictions, we need to go to very large system size, i.e., of order $N\sim 10^6$. 

\begin{figure}
    \centering
    \includegraphics[width=0.7\linewidth]{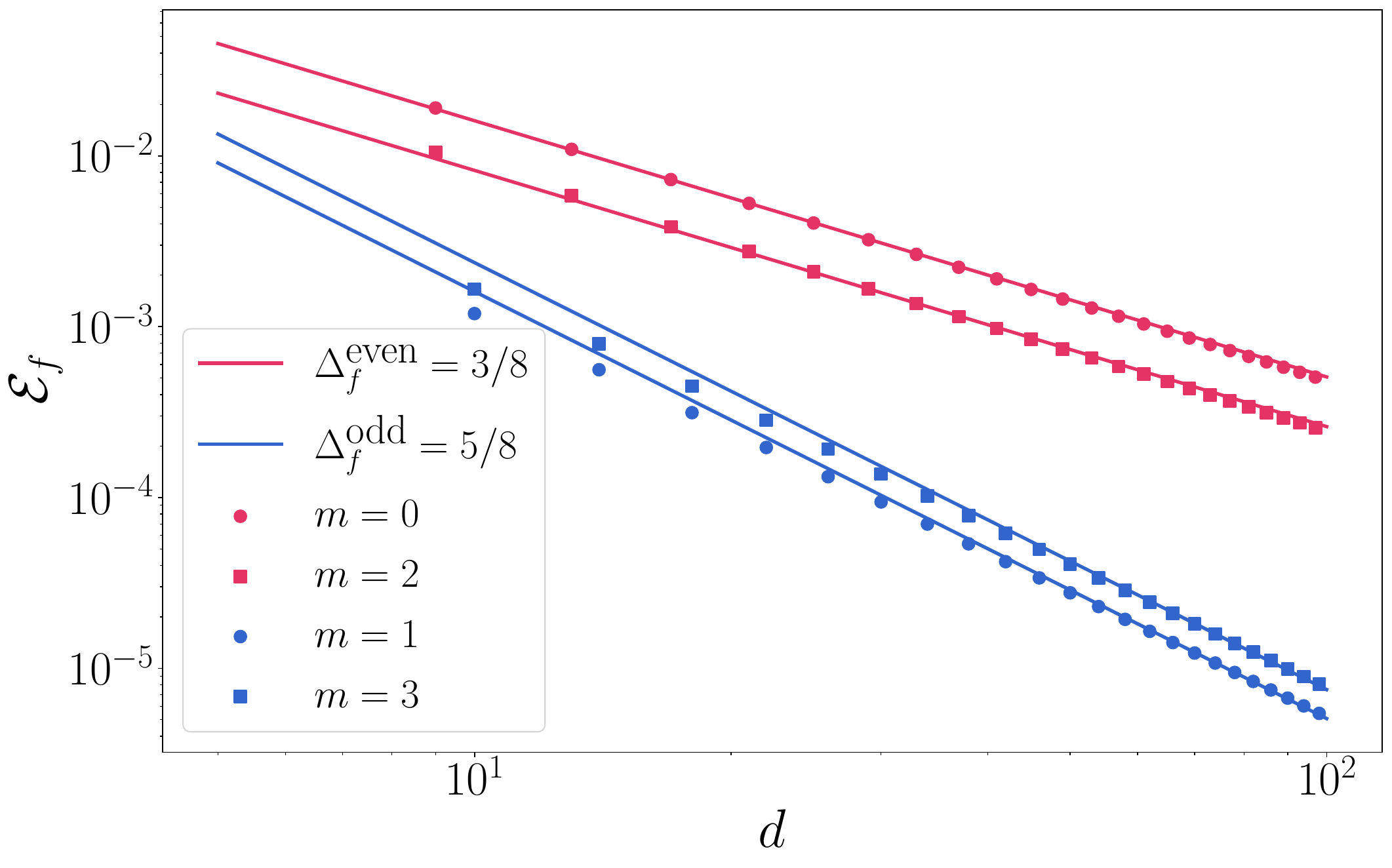}
    \caption{Boundary logarithmic negativity between sites $m$ and $m+d$ as a function of $d$ in log-log scale, for $m=0,1,2,3$ with $N=1000$ and $\rho=p=1/2$. The symbols are numerical calculations of the negativity, and the solid lines are either obtained from Eq.~\eqref{eq:C01d} for $m=0,1$ or numerical fit for $m=2,3$. The slope of the curves depends on the parity of $m$, and we have $\E_f \propto d^{-4\Delta_f}$ with $\Delta_f^{\textrm{even}}=3/8$ and $\Delta_f^{\textrm{odd}}=5/8$.}
    \label{fig:Efbndr}
\end{figure}

Based on Eqs.~\eqref{eq:C01d}, we thus see that the boundary negativity decays as $d^{-4\Delta_f}$, with $\Delta_f=3/8$ and $\Delta_f=5/8$ when the left site is at $m=0$ and $m=1$, respectively. We verified this numerically in Fig.~\ref{fig:Efbndr}. This is surprising for several reasons. First, the power-law decay is different from the bulk result $\Delta_f=1/2$. Second, one can easily check that the boundary negativity in the open homogeneous chain also decays with exponent $\Delta_f=1/2$. This indicates that, contrarily to what happens in the bulk, the Krawtchouk chain displays different physical behaviour close to the boundary compared to the homogeneous chain. Finally, the fact that the exponent depends on the position of the left site $m$ is rather puzzling. We investigated different boundary correlations $C_{m,m+d}$ for small $m$, and observed a parity effect: for even $m$ we have $\Delta_f^{\textrm{even}}=3/8$, whereas for odd $m$ we have $\Delta_f^{\textrm{odd}}=5/8$. The coefficients however are no longer given by Eq.~\eqref{eq:C01d} for $m>1$, and an analytical derivation for arbitrary $m$ remains to be found. We illustrate this parity effect for the logarithmic negativity in Fig.~\ref{fig:Efbndr} with $m=0,1,2,3$. 

\section{Conclusion}\label{sec:ccl}

In this paper, we have initiated the study of the fermionic logarithmic negativity in inhomogeneous free-fermion models. We focused on a one-dimensional model whose diagonalization relies on Krawtchouk polynomials, that are at the bottom of the discrete part of the Askey scheme for $q=1$. Previous studies on the entanglement entropy in this model have shown that it is described by a CFT with central charge $c=1$ in the scaling limit. We confirmed this by showing that the logarithmic negativity of adjacent intervals scales as $\E_f~\sim~\frac c4 \log (\ell_1\ell_2/(\ell_1+\ell_2))$ with $c=1$, for different choices of the ratio $\ell_1/\ell_2$. 

We also examined the decay of the logarithmic negativity for disjoint systems as a function of their lattice separation $d$. We concentrated our attention on the skeletal regime $\ell_1=\ell_2=1$, as it is known to be sufficient to extract the leading behaviour, and considered bulk and boundary negativities. In the bulk, the fermionic logarithmic negativity was found to decay as $\E_f \propto d^{-4\Delta_f}$ with $\Delta_f=1/2$. We were able to prove this result for (i) arbitrary $p$ and small filling fraction $\rho\ll 1$, and (ii) arbitrary $\rho$ and $p=1/2$. Combining these two results, we offered with Eq.~\eqref{eq:CGenCOnj} a conjecture for generic $p$ and~$\rho$ regarding the expression of the correlation function, and hence the skeletal logarithmic negativity. The exponent $\Delta_f=1/2$ corresponds to free Dirac fermions in one dimension, and agrees with previous results for the homogeneous free-fermion model. These results indicate that the inhomogeneities of the Krawtchouk chain do not play a significant role in the bulk, and that the entanglement properties of the model are identical to those of homogeneous free-fermions far from the boundaries. Close to the boundaries however, the story is different. We found that the negativity between sites $m$ and $m+d$ for small $m$ (i.e., close to the left boundary) decays as $\E_f \propto d^{-4\Delta_f}$ with an exponent that depends on the parity of $m$. For even $m$ we have $\Delta_f^{\textrm{even}}=3/8$, whereas for odd $m$ it is $\Delta_f^{\textrm{odd}}=5/8$. We were able to prove this behaviour for $m=0,1$, and provided numerical evidences for $m=2,3$. 

This work opens several natural research avenues. First, it would be important to explore the behaviour of the negativity in other free-fermion chains of the Askey scheme, such as the dual Hahn and Racah chains \cite{Crampe:2019upj}. In particular, one should investigate the boundary negativity, and determine whether the parity effect observed in the Krawtchouk chain reflects at a more general property of these models. Second, it is worth noting that the behaviour of the boundary correlations represents a rare instance where a physical property of the Krawtchouk chain strongly differs from its homogeneous counterpart (see also \cite{bernard2024currents}). It is thus an intriguing open question to understand the physical reason behind the peculiar behaviour of the boundary correlations observed here. Finally, the problem of the negativity in inhomogeneous setting should be explored in different contexts, such as nonequilibrium situations and curved-space CFTs \cite{DSVC17}. This would undoubtedly enhance our understanding of the role inhomogeneities play in realistic condensed-matter and cold-atomic systems.

\section*{Acknowledgements}
GP held FRQNT and CRM-ISM postdoctoral fellowships, and received support from the Mathematical Physics Laboratory of the CRM while this work was carried out. The research of LV is funded in part by a Discovery Grant from the Natural Sciences and Engineering Research Council (NSERC) of Canada. We thank Cl\'ement Berthiere for valuable comments on the draft. 

\appendix
\section{Asymptotics for the correlation functions}\label{sec:app}

In this appendix, we compute various asymptotic expansions for correlation functions of the Krawtchouk chain. We shall use the following asymptotic limit relating Krawtchouk and Hermite polynomials \cite{koekoek2010hypergeometric},
\begin{equation}
\label{eq:KtoH}
    \lim_{N \to \infty}\sqrt{\begin{pmatrix}
        N \\ k
    \end{pmatrix}} K_k(pN + x \sqrt{2p(1-p)N},p,N)= \frac{(-1)^k H_k(x)}{\sqrt{2^k k! \Big(\frac{p}{1-p}\Big)^k}},
\end{equation}
where $ H_k(x)$ is the Hermite polynomial of order $k$ evaluated in $x$. In this relation, $k$ is a constant. 

\subsection{Bulk correlation functions}
\label{sec:bulkC}

\subsubsection{Bulk correlation function for small filling fraction $\rho \ll 1$}

We compute below the correlation function corresponding to two sites centered around $pN$ (in the bulk of the chain) and separated by a distance $d$, 
as illustrated in Fig.~\ref{fig:bulkCmn}. 
Moreover, we consider the limit of small filling fraction, $\rho \ll 1$, and large system size, $N\to \infty$. 
We combine the exact form \eqref{eq:CmnExact} for the correlation function, the asymptotic formula \eqref{eq:KtoH} and the Christoffel-Darboux formula for Hermite polynomials, and find
\begin{equation}
    C_{pN-\frac d2,pN+\frac d2} = F\Big(pN-\frac d2,pN+\frac d2\Big) \frac{1}{K!2^{K+1}} \frac{H_K(-x) H_{K+1}(x)-H_K(x) H_{K+1}(-x)}{2x}
\end{equation}
where $x = -\frac{d}{\sqrt{8p(1-p)N}}$ and
\begin{equation}
    F(m,n) = (-1)^{m+n}\left(p^{m+n}(1-p)^{2N-m-n} \binom{N}{m} \binom{N}{n}\right)^{\frac 12}.
\end{equation}
Note that we used the asymptotic relation of Eq.~\eqref{eq:KtoH} with $k\sim K$, and that $K\sim \rho N$ is divergent in the large-$N$ limit, while Eq.~\eqref{eq:KtoH} is valid for constant $k$. To circumvent this issue, we assume in the following that the filling fraction is very small, $\rho\ll 1$. 

In the limit of large $k$, we have 
\begin{equation}
    \eE^{-x^2/2} H_{k}(x)\sim \frac{2^{k}}{\sqrt{\pi}} \Gamma\Big(\frac{k+1}{2}\Big) \cos \Big(x\sqrt{2k}-\frac{k\pi}{2} \Big),
\end{equation}
and hence
\begin{multline}
    C_{pN-\frac d2,pN+\frac d2}  \sim F\Big(pN-\frac d2,pN+\frac d2\Big) \frac{\eE^{x^2}}{\sqrt{\pi}2x} \Big( \cos \Big(x \sqrt{2K}+\frac{K\pi}{2}\Big) \cos\Big(x \sqrt{2K+2}-\frac{(K+1)\pi}{2}\Big) \\-\cos \Big(x \sqrt{2K+2}+\frac{(K+1)\pi}{2}\Big) \cos\Big(x \sqrt{2K}-\frac{K\pi}{2}\Big)\Big).
    \label{eq:CAsymptstep}
\end{multline}

 To proceed, we use Stirling's approximation to extract the asymptotic behaviour of the function $F(m,n)$. We find 
 \begin{equation}
     F\Big(pN-\frac d2,pN+\frac d2\Big) \sim \left( \frac{1}{2\pi p(1-p)N }\right)^{\frac 12} \eE^{-x^2}.
 \end{equation}
 We combine this result with Eq.~\eqref{eq:CAsymptstep}, and after some straightforward trigonometric calculations, we arrive at
 \begin{equation}
    C_{pN-\frac d2,pN+\frac d2 } \sim \frac{1}{\pi d}\sin\left( d \sqrt{\frac{\rho}{p(1-p)}} \right)
    \label{eq:CSmallFFAnalytical2}
\end{equation}
in the limit $N\to \infty$, which is Eq.~\eqref{eq:CSmallFFAnalytical}. 

\subsubsection{Bulk correlation function for $p=\frac{1}{2}$}\label{sec:Cphalf}
The correlation function of two sites separated by a distance $d$ and centered around the middle of the chain with $p = 1/2$ reads
\begin{multline}
        C_{\frac{N-d}{2},\frac{N+d}{2}} = (-1)^{N}  2^{-(N+1)} \sqrt{ \binom{N}{\frac{N-d}{2}} \binom{N}{\frac{N+d}{2}}} \binom{N}{K} \left( N - K \right) \\[0.3mm] \frac{ K_K \left(\frac{N+d}{2},\frac{1}{2},N \right) K_{K+1} \left(\frac{N-d}{2},\frac{1}{2},N \right) - K_K \left(\frac{N-d}{2},\frac{1}{2},N \right) K_{K+1} \left(\frac{N+d}{2},\frac{1}{2},N\right)}{d},
\end{multline}
where we impose that $(N+d)$ is an even number.
The identity 
\begin{equation}
    K_n \left(N-x,\frac{1}{2}, N \right) = (-1)^{n} K_n \left(x,\frac{1}{2}, N \right)
\end{equation}
allows to recast the correlation function as 
\begin{equation}
            C_{\frac{N-d}{2},\frac{N+d}{2}} = (-1)^{N+K}  2^{-N} \sqrt{ \binom{N}{\frac{N-d}{2}} \binom{N}{\frac{N+d}{2}}} \binom{N}{K} \left( N - K \right) \\[0.3mm] \frac{ K_K \left(\frac{N-d}{2},\frac{1}{2},N \right) K_{K+1} \left(\frac{N-d}{2},\frac{1}{2},N \right)}{d}.
\end{equation}

Our goal is to compute the asymptotic behaviour of this correlation function when $N \rightarrow \infty$. In Ref.~\cite{ismail1998strong}, an asymptotic expansion of the Krawtchouk polynomials $P_n(nt,\frac{1}{2},N) \equiv (\frac{-1}{2})^{n} \binom{N}{n}K_{n}(nt,\frac{1}{2},N)$ is derived for the special case where the ratio $N/n$ is constant as $n \rightarrow \infty$ and $N \rightarrow \infty$. In terms of the polynomials $P_n(nt,\frac{1}{2},N)$, the correlation function reads
\begin{equation}
\label{eq:corr P}
            C_{\frac{N-d}{2},\frac{N+d}{2}} = (-1)^{N+K+1}  2^{-N+2K+1} \left( N - K \right)\frac{\sqrt{ \binom{N}{\frac{N-d}{2}} \binom{N}{\frac{N+d}{2}}}}{\binom{N}{K+1}}  \frac{ P_K \left(t_1 K,\frac{1}{2},N \right) P_{K+1} \left(t_2(K+1),\frac{1}{2},N \right)}{d}
\end{equation}
where we introduced
$t_1 = \frac{N-d}{2 K}$ and $t_2 = \frac{N-d}{2 (K+1)}$. The asymptotic expansion of the Krawtchouk polynomials is\footnote{We added a factor $2 \times (-1)^{\frac{N+d}{2}+K+1}$ compared to Eq.~(3.8) of \cite{ismail1998strong}.} \cite{ismail1998strong}
\begin{equation}
     P_n( nt, \frac{1}{2}, N) = (-1)^{\frac{N+d}{2}+K+1} \frac{1}{2^{n-1} \pi} \textrm{e}^{-n \Re\{p(t)\}} \sin\left( n \alpha(t) - \alpha_1(t)\right) \left| \frac{\sqrt{\pi} n^{-\frac{1}{2}}}{z_1(t)(2p''(t))^{\frac{1}{2}}} \right|,
\end{equation}
where the function $p(t)$ is defined as 
\begin{equation}
    p(t) = \log(z_1(t)) - t \log(1+z_1(t)) - (\gamma -t) \log(1-z_1(t)),
\end{equation}
with
\begin{align}
    z_1(t)&=\frac{-\Delta(t) + \sqrt{D(t)}}{2 ( \gamma-1)},\\
    \Delta(t) &= \gamma - 2t, \\
    D(t) &= \Delta(t)^2 - 4 ( \gamma - 1)\\
    \gamma&=\frac{N}{n}.
\end{align}
Moreover, we have
\begin{equation}
    p''(t) = \frac{2(\gamma - 1) z_1(t) + \Delta(t)}{z_1(t)(1-z_1(t)^2)}.
\end{equation}
Looking at Eq.~\eqref{eq:corr P}, we see that the two relevant $\gamma$'s for this problem are $\gamma_1 = \frac{N}{\rho(N+1) -1}$ and $\gamma_2 = \frac{N}{\rho (N+1)}$, where we recall $K = \rho (N + 1) -1$. They are both constant in the large-$N$ limit, as required. 
The remaining functions that need to be defined are
\begin{equation}
    \alpha(t) = t \cot^{-1} \Big( \frac{2(\gamma - 1) - \Delta(t)}{\sqrt{-D(t)}}\Big) + (\gamma - t)\left(\cot^{-1}\Big( \frac{2(\gamma - 1) + \Delta(t)}{-\sqrt{-D(t)}}\Big) - \pi\right) - \cot^{-1}\Big(\frac{-\Delta(t)}{\sqrt{-D(t)}}\Big),
\end{equation}
and
\begin{equation}
    \alpha_1(t) =  \textrm{arg}\left\{ z_1(t)  p''(t)^{\frac{1}{2}}\right\}.
\end{equation}
With these expressions in hand, the correlation function can be recast as 
\begin{multline}
\label{eq:corr asympt}
     C_{\frac{N-d}{2},\frac{N+d}{2}} = \frac{(-1)^{N+K}  2^{-N+2} \left( N - K \right)}{d \pi}\frac{\sqrt{ \binom{N}{\frac{N-d}{2}} \binom{N}{\frac{N+d}{2}}}}{\binom{N}{K+1}} 
    \textrm{e}^{-K \Re\{p(t_1)\}-(K+1)\Re\{p(t_2)\}} \\
    \times  \sin\left(K \alpha(t_1) - \alpha_1(t_1)\right)   
       \sin\left((K+1) \alpha(t_2) - \alpha_1(t_2)\right) \left| \frac{ K^{-\frac{1}{2}}}{z_1(t_1)(2p''(t_1))^{\frac{1}{2}}} \right|\left| \frac{(K+1)^{-\frac{1}{2}}}{z_1(t_2)(2p''(t_2))^{\frac{1}{2}}} \right|.
\end{multline}

Now, let us first examine the product of the two sines in Eq.~\eqref{eq:corr asympt}. In terms of the variables $d$, $\rho$ and $N$, after a few simplifications, this product reads
\begin{multline}
\label{eq:sins}
    \sin\left( K\alpha(t_1) - \alpha_1(t_1)\right)  \sin\left( (K+1)\alpha(t_2) - \alpha_1(t_2)\right) =\cos \left[ \frac{1}{2} \left( \pi + d \pi + N \pi - 2 \rho \varphi_1  + N \left( \varphi_3 - \varphi_2 - 2 \rho \varphi_1  \right) \right) \right.\\[0.3cm]  \left.
    + \frac{d}{2} \left( \varphi_2 + \varphi_3 \right) +  \textrm{arg}\left\{(\textrm{i}\zeta_1 - d) \left( \frac{\textrm{i} \zeta_1(N (1 - \rho ) + \rho)}{(d -\textrm{i}\zeta_1)(d^2 + 2 N (N (\rho - 1) + \rho) - \textrm{i} d \zeta_1)} \right)^{\frac{1}{2}} \right\}  \right] \\[0.3cm]
 \times \cos\left[  \frac{1}{2} \left(  \pi + d  \pi + N  \pi + 2( 1- \rho ) \varphi_4 + N ( \varphi_5 + \varphi_6 - 2 \rho \varphi_4 ) + d ( \varphi_5 - \varphi_6)\right)  \right. \\[0.3cm]  \left. 
 +  \textrm{arg}\left\{ ( \textrm{i}\zeta_2 - d ) \left(\frac{\textrm{i}\zeta_2}{(d-\textrm{i}  \zeta_2 )  (d^2 +
   2 N (N+1)  (\rho - 1) - \textrm{i} d \zeta_2)}  \right)^{\frac{1}{2}} \right \} \right]
\end{multline}
where 
\begin{equation*}
    \zeta_1 =  \sqrt{-d^2 - 4 (N+1) \rho (N (\rho - 1) + \rho)}
,\quad \zeta_2 = \sqrt{-d^2 - 4 (N+1) (\rho - 1) (\rho -1 + N \rho)},
\end{equation*}
and
\begin{align}
\nonumber
  \hspace{-.4cm}  &\varphi_1 =\cot^{-1} \left(\frac{d}{\zeta_1}\right), \hspace{.4cm}  &&\varphi_2 = \cot^{-1}\left( \frac{-d -2(N(\rho - 1) + \rho)}{\zeta_1}\right),\\ \nonumber
   & \varphi_3 = \cot^{-1}\left( \frac{d -2(N(\rho - 1) + \rho)}{\zeta_1}\right),  &&\varphi_4 = \cot^{-1} \left(\frac{d}{\zeta_2}\right), \\ \nonumber
    &\varphi_5 = \cot^{-1}\left( \frac{d - 2 (N+1)  (\rho - 1) }{\zeta_2}\right),   &&\varphi_6 = \cot^{-1}\left( \frac{d + 2 (N+1)  (\rho - 1) }{\zeta_2}\right).
\end{align}
%
%
%
%
The series expansions of the terms in the cosines in Eq.~\eqref{eq:sins}  
are
\begin{align}
    \varphi_1 &= \frac{\pi}{2} + \mathcal{O}(N^{-1}), \\
    \varphi_3- \varphi_2 - 2\rho \varphi_1 &= -\rho \pi + \mathcal{O}(N^{-2}), \\
    \varphi_2 + \varphi_3 & = 2\arcsin(\sqrt{\rho})+\mathcal{O}(N^{-1}),
\end{align}
and 
\begin{align}
    \varphi_4 &= \frac{\pi}{2} + \mathcal{O}(N^{-1}), \\
   \varphi_5 + \varphi_6 - 2 \rho \varphi_4 &= -\rho \pi +\mathcal{O}(N^{-1}) ,\\
    \varphi_5 - \varphi_6 &= 2 \arcsin(\sqrt{\rho}) + \mathcal{O}(N^{-1}).
\end{align}
Moreover, the $\textrm{arg}$ functions in both cosines go to $\pi/2$. Putting this all together gives
%
%
%
\begin{equation}
\label{eq:sin serie}
    \sin\left( K\alpha(t_1) - \alpha_1(t_1)\right)   
       \sin\left( (K+1)\alpha(t_2) - \alpha_1(t_2)\right) \sim \frac 12 (-1)^{(N+1) \rho + 1} \sin(2 d \arcsin(\sqrt{\rho})).
\end{equation}

The second step is to find the behaviour of the binomial terms in Eq.~\eqref{eq:corr asympt} as $N \rightarrow \infty$. The asymptotic expansion reads
%
%
%
\begin{equation}
\label{eq:bin serie}
    \frac{\sqrt{ \binom{N}{\frac{N-d}{2}} \binom{N}{\frac{N+d}{2}}}}{\binom{N}{K+1}}  \sim \exp\left( \frac{1}{2 N} + \frac{1}{12N \rho (1-\rho)}\right)  \frac{2^N(6 N- 3 d^2 -2 )}{3 N}(\rho^{-1} -1)^{-(N+1) \rho}  (1 - \rho)^N  \sqrt{(1 - \rho) \rho}.
\end{equation}

Turn then to the exponential term in Eq.~\eqref{eq:corr asympt}. 
%
%
After some algebra, it can be recast as
\begin{multline}
    \exp\left(-K \Re\{p(t_1)\}-(K+1)\Re\{p(t_2)\} \right ) = 
    \\[0.3cm]
    \left|  \left ( \frac{d - \textrm{i} \zeta_1}{2 (N (\rho - 1) + \rho)} \right)^{- (N+1)  \rho} 
    \left( \frac{d - 2 N + 2 \rho + 2 N \rho - \textrm{i} \zeta_1}{2 (N (\rho - 1) + \rho)}\right)^{\frac{N-d}{2}} 
    \left( \frac{-d - 2 N + 2 \rho + 2 N \rho + \textrm{i} \zeta_1}{2 (N (\rho - 1) + \rho)}\right)^{\frac{N + d}{2}}
    \right. \\[0.3cm] \left. 
    \left( \frac{d - \textrm{i} \zeta_2}{2 (N+1) (\rho - 1)}\right)^{1 -(N+1)  \rho} \left( -\frac{d - 2 (N+1) (\rho - 1) - \textrm{i}\zeta_2}{2 (N+1) (\rho - 1)} \right)^{\frac{N + d}{2}} 
    \left( -\frac{-d - 2 (N+1) (\rho - 1) + \textrm{i} \zeta_2}{2 (N+1) (\rho - 1)} \right)^{\frac{N - d}{2}} \right|,
\end{multline}
and the asymptotic expansion of this expression is 
\begin{equation}
\label{eq:exp serie}
    \exp\left(-K \Re\{p(t_1)\}-(K+1)\Re\{p(t_2)\} \right ) \sim \exp \left( \frac{2 + 2  d^2 +\frac{1}{(\rho - 1) \rho}}{4N}\right) (1 - \rho)^{-\frac{1}{2} + N (\rho - 1) + \rho}  \rho^{\frac{1}{2} - (N+1) \rho}.
\end{equation}

%
%
Finally, the asymptotic expansion of the remaining terms in Eq.~\eqref{eq:corr asympt} is
\begin{equation}
\label{eq:abs serie}
    \left| \frac{(K+1)^{-\frac{1}{2}}}{z_1(t_2)(2p''(t_2))^{\frac{1}{2}}} \right| \left| \frac{ K^{-\frac{1}{2}}}{z_1(t_1)(2p''(t_1))^{\frac{1}{2}}} \right| \sim \frac{1}{8}\sqrt{\frac{-d^2 (1 - 2 \rho)^2 (\rho - 1) \rho + (1 - 
   2 (N - 2) (\rho - 1) \rho)^2}{N^4 (\rho - 1)^4 \rho^4}}.
\end{equation}

Assembling Eqs.~\eqref{eq:corr asympt}, \eqref{eq:sin serie}, \eqref{eq:bin serie}, \eqref{eq:exp serie} and \eqref{eq:abs serie} gives
\begin{multline}
    C_{\frac{N-d}{2}, \frac{N+d}{2}} \sim (-1)^{N+1}\exp \left( \frac{1}{N} + \frac{d^2}{2 N} + \frac{1}{6 N \rho (\rho -1)}\right) (N + 1) (6 N - 2 - 3 d^2)\\[0.3cm] 
   \times  \frac{\sqrt{-d^2 (1 - 2 \rho )^2 (\rho - 1) \rho + (1 - 
   2 (N - 2) (\rho - 1) \rho)^2}}{12 d N^3 \pi (\rho - 1) \rho} \sin \left( 2 d \arcsin(\sqrt{\rho})\right).
\end{multline}
A final series expansion for $N \rightarrow \infty$ with even $N$ yields 
\begin{equation}
    C_{\frac{N-d}{2}, \frac{N+d}{2}} \sim \frac{1}{\pi d} \sin \left(2 d \arcsin(\sqrt{\rho}) \right),
\label{eq:CrhoHalfApp}
\end{equation}
which is Eq.~\eqref{eq:CRhoPhalf}. The same limit for odd $N$ flips the sign in Eq.~\eqref{eq:CrhoHalfApp}, but this does not affect the behaviour of the logarithmic negativity. Therefore, we focus on even $N$ for simplicity.

\subsection{Boundary correlation functions}\label{sec:bndrC}

Here, we compute the correlation function between the left boundary ($n=0$) of the chain, and a site at distance $d$, as illustrated in Fig.~\ref{fig:bndrCmn}. The Krawtchouk polynomials evaluated at $n=0$ is $K_k(0,p,N)=1$, and therefore the correlation function in Eq.~\eqref{eq:CmnCD} reads 
\begin{equation}
    C_{0,d} = (-1)^d \Big(\frac{p}{1-p}\Big)^{d/2} (1-p)^N \sqrt{\binom{N}{d}} \sum_{k=0}^K  \Big(\frac{p}{1-p}\Big)^{k}\binom{N}{k}K_d(k,p,N),
    \label{eq:C0dEx}
\end{equation}
where we used the symmetry of the Krawtchouk polynomials $K_{k}(d,p,N) = K_{d}(k,p,N)$. To proceed, we use Rodrigues' formula for Krawtchouk polynomials \cite{koekoek2010hypergeometric}, 
\begin{equation}
    \Big(\frac{p}{1-p}\Big)^{k}\binom{N}{k}K_d(k,p,N) = \nabla^d \left( \binom{N-d}{k}  \Big(\frac{p}{1-p}\Big)^{k}\right),
    \label{eq:RF}
\end{equation}
where $\nabla f(k) \equiv f(k+1)-f(k)$ is the discrete derivative. With its help, Eq.~\eqref{eq:C0dEx} becomes a telescopic sum. We use Eq.~\eqref{eq:RF} a second time to recast the result in terms of Krawtchouk polynomials. We find, up to a negligible term,
\begin{equation}
     C_{0,d} = (-1)^d \Big(\frac{p}{1-p}\Big)^{\frac d2+K} (1-p)^N \sqrt{\binom{N}{d}} \binom{N-1}{K} K_{d-1}(K,p,N-1) .
\end{equation}
For simplicity, we now impose $K=p(N-1)$, which corresponds to $\rho \sim p$ in the large-$N$ limit. With this choice of $K$, we use the asymptotic relation of Eq.~\eqref{eq:KtoH} with $x=0$. Given that the Hermite polynomials evaluated in $x=0$ satisfy $H_{d-1}(0) = 2^{d-1}\sqrt{\pi}\Gamma(1-d/2)^{-1}$, we find
\begin{equation}
     C_{0,d} = (-1)\Big(\frac{p}{1-p}\Big)^{\frac 12+K} (1-p)^N\sqrt{\frac{\binom{N}{d}}{\binom{N-1}{d-1}}}\binom{N-1}{p(N-1)}\left(\frac{2^{d-1}\pi}{(d-1)!}\right)^{\frac 12}\frac{1}{\Gamma\Big(1-\frac d2\Big)}.
\end{equation}
Using Stirling's approximation for the binomial coefficients, we further simplify $C_{0,d}$ to 
\begin{equation}
    C_{0,d}\sim(-1)\frac{1}{(2d)^{\frac 12}}\left(\frac{2^{d-1}}{(d-1)!}\right)^{\frac 12}\frac{1}{\Gamma\Big(1-\frac d2\Big)}.
\end{equation}
In the limit of large $d$, the leading order is 
\begin{equation}
    C_{0,d} \sim(-1)\frac{1}{(2\pi^3)^{\frac 14}d^{\frac 34}}\sin\Big(\frac{\pi d}{2}\Big), 
\end{equation}
where the sine function arises because the Gamma function has poles for negative integers. This result is precisely Eq.~\eqref{eq:C0d}. 

The calculation for $C_{1,d+1}$ follows a similar reasoning. One has $K_{1}(k,p,N) = 1-\frac{k}{pN}$. The sum in Eq.~\eqref{eq:CmnCD} thus becomes a double sum, both of which can be reduced to single Krawtchouk polynomials using Rodrigues' formula. Alternatively, one can directly start from the expression of Eq.~\eqref{eq:CmnCD} with $(m,n)=(1,d+1)$. Both approaches yield the same results, even though it is not manifest at first sight. To show the equivalence between the two results, it is useful to be reminded of the identity \cite{crampe2023lambda}
\begin{equation}
   K_k(x-1, N-1,p) = \frac{pN}{x}( K_k(x, N,p) -  K_{k+1}(x, N,p)  ).
\end{equation}
The asymptotic calculations are then similar to those for $C_{0,d}$ outlined above, but are more cumbersome. 

\providecommand{\href}[2]{#2}\begingroup\raggedright\endgroup

\end{document}